\documentclass[11pt]{article}
\usepackage{xspace}
\usepackage{graphicx}
\usepackage{amsmath}
\usepackage{amssymb}
\usepackage{color}

\usepackage{subfig}

\definecolor{Red}{rgb}{1,0,0}
\definecolor{Green}{rgb}{0,1,0}
\definecolor{Blue}{rgb}{0,0,1}
\definecolor{Black}{rgb}{0,0,0}

\def\beq{\begin{equation}}
\def\eeq#1{\label{#1}\end{equation}}
\def\eeqn{\end{equation}}

\def\beqa{\begin{eqnarray}}
\def\eeqa#1{\label{#1}\end{eqnarray}}
\def\eeqan{\end{eqnarray}}

\let\bar=\overbar

\def\Dslash{\not{\hbox{\kern-4pt $D$}}}
\def\dslash{\not{\hbox{\kern-2pt $\del$}}}

\def\msb{{\bar{\ssstyle M \kern -1pt S}}}

\def\Title#1{\begin{center} {\Large {\bf #1} } \end{center}}

\begin{document}

\Title{Study of the intrinsic electron neutrino component in the T2K neutrino beam with the near detector, ND280}

\bigskip\bigskip


\begin{raggedright}  

{\it Sophie E King on behalf of the T2K collaboration\index{King, S.},\\
School of Physics and Astronomy\\
Queen Mary University of London\\
E1 4NS London, UK}\\

\end{raggedright}
\vspace{1.cm}

{\small
\begin{flushleft}
\emph{To appear in the proceedings of the Prospects in Neutrino Physics Conference, 15 -- 17 December, 2014, held at Queen Mary University of London, UK.}
\end{flushleft}
}

\section{T2K and electron neutrinos}

T2K (Tokai to Kamioka) is a long baseline neutrino oscillation experiment optimised to measure $\theta_{13}$ through electron neutrino appearance in a muon neutrino beam.  This is achieved by coinciding the beam peak energy, $\sim0.6$ GeV, with the first $\nu_e$ appearance probability maximum.  The far detector, Super-Kamiokande (SK), is a Cherenkov light detector situated 295km `downstream' of the accelerator.  T2K adopts an off-axis setup to narrow and increase the flux energy spectrum peak, with SK at an angle $2.5^{\circ}$ with respect to the beam.  Positioned 280m from the neutrino production target, along the same axis as SK, is the near detector, ND280, which characterises the beam pre-(standard 3 neutrino) oscillation.  

For $\nu_{\mu} \rightarrow \nu_e$ oscillation searches, the signal at SK is electron neutrinos.  The biggest background comes from the intrinsic $\nu_e$ component of the beam itself, due largely to the decay of muons in the decay tunnel \cite{t2knueapp2012}.  The precision with which the contamination is modelled, and our poor understanding of $\nu_e$ cross-sections, therefore limit the uncertainty.

Here we focus on the {\it$\nu_e$ Tracker Analysis} which identifies Charged Current (CC) electron neutrino events at ND280.  This selection of $\nu_e$ interactions may be used to check the Monte Carlo (MC) predicted beam contamination and to perform cross-section measurements.

\section{ND280}

\begin{figure}[!ht]
\begin{center}
\begin{minipage}{0.48\textwidth}
\includegraphics[width=1\columnwidth]{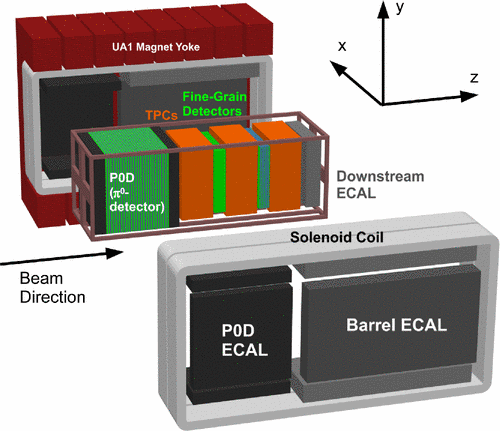}
\caption{Schematic of ND280}
\label{fig:nd280}
\end{minipage}
\hspace{24pt}
\begin{minipage}{0.38\textwidth}
\includegraphics[width=1\columnwidth]{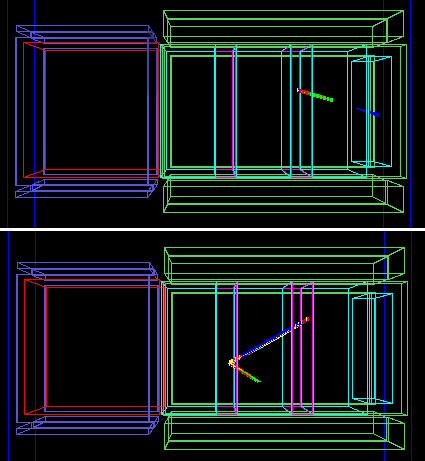}
\caption{ND280 displays of MC $\nu_e$ CC0$\pi$ interactions.  Top: $e^-$ reconstructed.  Bottom: $e^-$ and $p^+$. }
\label{fig:events}
\end{minipage}
\end{center}
\end{figure}

Made up of multiple sub-detectors grouped inside a magnet, ND280 is designed to constrain SK flux and cross-section model parameters; this significantly reduces the systematic error on T2K oscillation results when combined with the SK measurements.  ND280 is depicted in Figure~\ref{fig:nd280}, where `downstream' (`upstream') is defined as the +z (-z) direction.  The target mass is provided by two Fine Grained Detectors (FGDs), namely FGD1 comprised of carbon scintillator layers, and FGD2 with alternating water and carbon layers.  These have good vertex reconstruction and their size is optimised such that a large fraction of particles have enough energy to travel through one of Time Projection Chambers (TPCs) and possibly the Electromagnetic Calorimeters (ECals).  TPC information is used to calculate the momentum and charge of tracks that travel in the magnetic field, furthermore the energy deposited as a function of distance gives excellent particle identification (PID) capabilities.  If a track enters an ECal then the distribution of charge may be used to perform further PID where track-like (muon) and shower-like (electron) objects are distinguished.  FGDs/TPCs are numbered in the downstream direction.  These three detectors form the `tracker'; upstream of these ND280 contains a $\pi^0$ detector (P$\emptyset$D) with separate ECals.

\vspace{-20pt}

\section{$\nu_e$ Tracker Analysis}

The presence and flavour of CC neutrino interactions is determined by detecting the resulting lepton.  The $\nu_e$ tracker analysis aims to identify electrons from CC $\nu_e$ events; these must be separated from a background of other particles, mainly muons from $\nu_{\mu}$ interactions.  Identifying an electron is necessary, but it is not sufficient; there are many other interactions that can produce an electron and we must impose further conditions to reject these events.  The most common is photon conversions via pair production, which we refer to as {\it $\gamma$ background}.  These photons convert inside the FGD, but they may result from neutrino interactions that originate inside or outside of the FGD.

\subsection{The selection process}

%
\textbf{Event Quality} - Perform data quality and timing compatibility checks. 
%
\newline\textbf{Track selection} - Select the highest momentum negative track that originates in FGD1 (FGD2) and enters TPC2 (TPC3).  Require the momentum to be greater than 200MeV/c.
%
\newline\textbf{TPC track Quality} - The track must have a large enough TPC segment.  This is to maintain good TPC PID capabilities.	
%
\newline\textbf{PID} - Impose PID criteria on the selected track.  Cuts are made on TPC and ECal PID variables to accept electron-like tracks, and reject muon-, pion- and proton-like tracks.  
%
\newline\textbf{Reject pair production events} - Look to reject $\gamma$ background by cutting on the invariant mass and distance between the selected track and any secondary track that is positive with an electron-like TPC track.
%
\newline\textbf{Upstream Vetoes}  -  Check for upstream activity in the P$\emptyset$D and ECals, and for upstream TPCs tracks; these indicate that the identified vertex may not be the neutrino vertex.
%
\paragraph{}
At this stage the selection is split into two samples.  These are defined according to their topology, i.e. the particles that exit the nucleus.
\begin{itemize}
\item $\nu_e$ CC0$\pi$ - No pions exit the nucleus.
\item $\nu_e$ CCother -  All $\nu_e$ CC events that are not $\nu_e$ CC0$\pi$
\end{itemize}

\paragraph{CC0$\pi$,  No Michel electrons} – There are no Michel electron candidates.
\newline\textbf{CC0$\pi$, 1 FGD/TPC track} - No `extra' tracks originating in the FGD.  The definition of an 'extra' track is explained in the following section.
\newline\textbf{CC0$\pi$, ECal Activity} - For FGD2 events, require no ECal objects except those associated with the selected track.
\newline\textbf{CCother, CCnonQE} - Require a Michel electron, or at least one `extra' track starting near the selected track.
%

\section{Recent improvements}

Several features, such as the counting of Michel electrons and FGD-only tracks, that were only implemented in FGD1 are now included in both since the determination of systematics.  This increases the CC0$\pi$ purity and CCother efficiency as expected.  The definition of `extra' tracks mentioned in the selection cuts changed.  Previously, an extra track was any that starts in the same FGD as the selected track.  This was designed to detect events where an electron is ejected from the nucleus and reconstructed in the FGD and TPC, as depicted in the top reconstructed event display of Figure~\ref{fig:events}.  However, due to final state interactions (FSI), a proton may be ejected and reach the TPC; in this case, if the proton is reconstructed it should not cause the event to fail the selection.  Therefore, an `extra' track in the TPC is defined as one that is not proton-like according to the TPC PID.  This significantly increases the $\nu_e$ CC0$\pi$ efficiency, while the purity undergoes no significant change.

\section{Final selections}
\begin{figure}[!ht]
\begin{center}
\includegraphics[width=0.80\columnwidth]{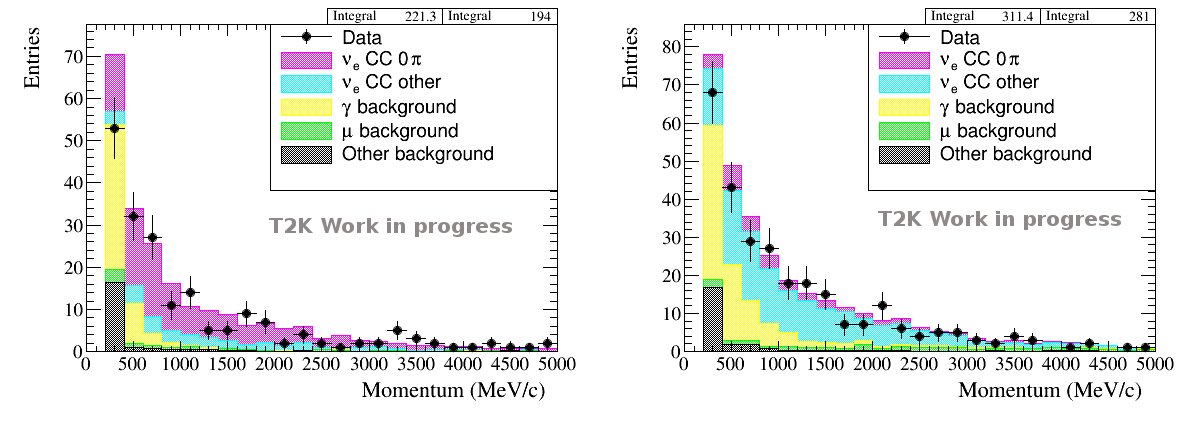}
\caption{Priliminary final selections for the $\nu_e$ CC0$\pi$ (left) and $\nu_e$ CCother (right) samples in (FGD1 + FGD2) as a function of the exiting electron momentum  }
\label{fig:finalSel}
\end{center}
\end{figure}

The combined FGD1 and FGD2 selections for the $\nu_e$ CC0$\pi$ (left) and $\nu_e$ CCother (right) samples are shown in Figure~\ref{fig:finalSel}.  The $\mu$ background refers to events where the selected track is a muon, and $\gamma$ background indicates a pair production interaction in the FGD.  The purity of the CC0$\pi$ and CCother samples are $49 ~ \%$ and $45 ~ \%$ respectively, with an efficiency each of $36.6 ~ \%$ and $29.7 ~ \%$.  Please note that these results are preliminary and do not yet include data taken 2014-2015.

\vspace{-20pt}

\section{Cross section plans}

Previously, T2K measured the $\nu_e$ CC inclusive (CC0$\pi$ + CCother) cross-section on FGD1 (carbon) \cite{ben_nueCC_xs}.  With increased data statistics (still accumulating at the time of writing) from recent data taking periods, the succeeding measurement is to be performed on the CC0$\pi$ sub-sample, which is of course more favourable for kinematic reconstruction.  The differential cross-section will be calculated as a function of electron momentum, angle and $Q^2$.  The biggest backgrounds are $\nu_e$ CCother and $\gamma$ events.  One possibility is to use the cut that rejects $\gamma$ events, which gives a very high purity sample ($>95~\%$) with reasonable statistics, to constrain the MC prediction; in this case the  $\nu_e$ CCother could be modified to contain less signal in order to form a sideband for this background.

\section{Summary}

The $\nu_e$ tracker analysis at ND280 detects $\nu_e$ CC interaction and splits into CC0$\pi$ and CCother topology based sub-samples, each with a purity approaching 50$~\%$.  Additional data will be added to the samples and $\nu_e$ CC0$\pi$ differential cross-section measurements on FGD1 (carbon) in electron variables is to be performed.

\section{Acknowledgments}

This work is presented on behalf of the T2K collaboration, and acknowledgement is given to the NExT/SEPnet Institute for financial support.


\begin{thebibliography}{99}



\bibitem{t2knueapp2012}
  K. Abe {\it et al.} (T2K Collaboration),
    Phys.\ Rev.\ Lett.\  {\bf 112} (2014) 061802.



\bibitem{ben_nueCC_xs}
  K. Abe {\it et al.} (T2K Collaboration),
  Phys.\ Rev.\ Lett.\  {\bf 113} (2014) 241803.


\end{thebibliography}
\end{document}